\documentclass{cimento}

\usepackage{graphicx}  

\title{Overview and performance of the 2023 MUGAST@LISE campaign at GANIL}
\author{V.~Girard-Alcindor\from{ins:x}\ETC,
H.~Jacob\from{ins:x}
M.~Assie\from{ins:x},
D.~Beaumel\from{ins:x}
Y.~Blumenfeld\from{ins:x}. On behalf of the MUGAST collaboration.}

\instlist{\inst{ins:x} Universit\'{e} Paris-Saclay, CNRS/IN2P3, IJCLab, 91405 Orsay, FranceIJCLAB
}

\begin{document}

\maketitle

\begin{abstract}
  MUGAST is a state-of-the-art silicon array combining trapezoidal and square shaped double-sided silicon strip detectors (DSSD) to four MUST2 telescopes. Coupled to a $\gamma$-ray spectrometer, the excellent angular coverage and compacity of the MUGAST array make it an ideal tool for the study of transfer reactions. It is a first step toward the development of the new generation of silicon arrays using pulse shape analysis (PSA) for particle identification, such as the future GRIT array developed by our collaboration. 
  \\

  In recent years, MUGAST has been widely used at GANIL. First with the AGATA $\gamma$-ray spectrometer and the VAMOS large acceptance spectrometer for the study of ISOL beams from the SPIRAL1 facility. It is now coupled with twelve EXOGAM clovers and to a new zero degree detection system at the end of the LISE fragmentation beamline.
\end{abstract}

\section{Introduction}

The study of direct-reactions in inverse kinematics is a powerful tool for investigating nuclear structure and nuclear astrophysics. Its highly selective reaction mechanism allows precise theoretical calculations which can reveal crucial informations about the structure of the nuclei. Transfer reactions are particularly suited for the study of shell-evolution, pairing, clustering and can also provide interesting insights on the reaction rates of astrophysically significant reactions. 

The development of new radioactive ion beams has revolutionized our understanding of the atomic nucleus and its structure. In the last 50 years an important effort has been made to measure and characterize the structure of exotic nuclei from the valley of stability toward the drip-lines. GANIL provides both ISOL type beam of up to 10 MeV/u with the SPIRAL 1 facility and fragmentation beam of $\sim$ 30-50 MeV/u using the LISE spectrometer.

The MUST2 and MUGAST array have taken advantage of both beam lines. Fragmentation beams were used in 2009 \cite{Giron, Burgunder} and 2014 \cite{Lecrom, Pereira} combining MUST2 telescopes to 4 EXOGAM clovers then from 2017 to 2018 \cite{Lalanne1, Lalanne2, Lalanne3, Valerian1, Valerian2} using the MUST2 telescopes coupled to a zero degree detection system for the detection of the recoils. From 2019 to 2021 \cite{Valerian1, Valerian2} four MUST2 telescopes were included in the MUGAST array and coupled to the AGATA $\gamma$-ray spectrometer in combination with the VAMOS large acceptance spectrometer. This so-called MUGAST-AGATA-VAMOS campaign was focused around the study of stripping reactions using $\sim$ 10 MeV/u ISOL beams delivered by GANIL's SPIRAL1 facility. Details of this campaign are described in Ref. \cite{MUGASTAGATAVAMOS}. 

Following the success of this first MUGAST campaign and to take advantage again of the LISE fragmentation beams, it was decided to perform a new MUGAST campaign. The experimental set-up is located at the end of the LISE spectrometer, and will run until the end of 2025. The MUGAST array will be combined to the CATS tracker, EXOGAM clovers and a new zero degree detection system (ZDD).

\section{Beam production}

\begin{figure}[ht]
  \centering
  \includegraphics[width=0.7\textwidth]{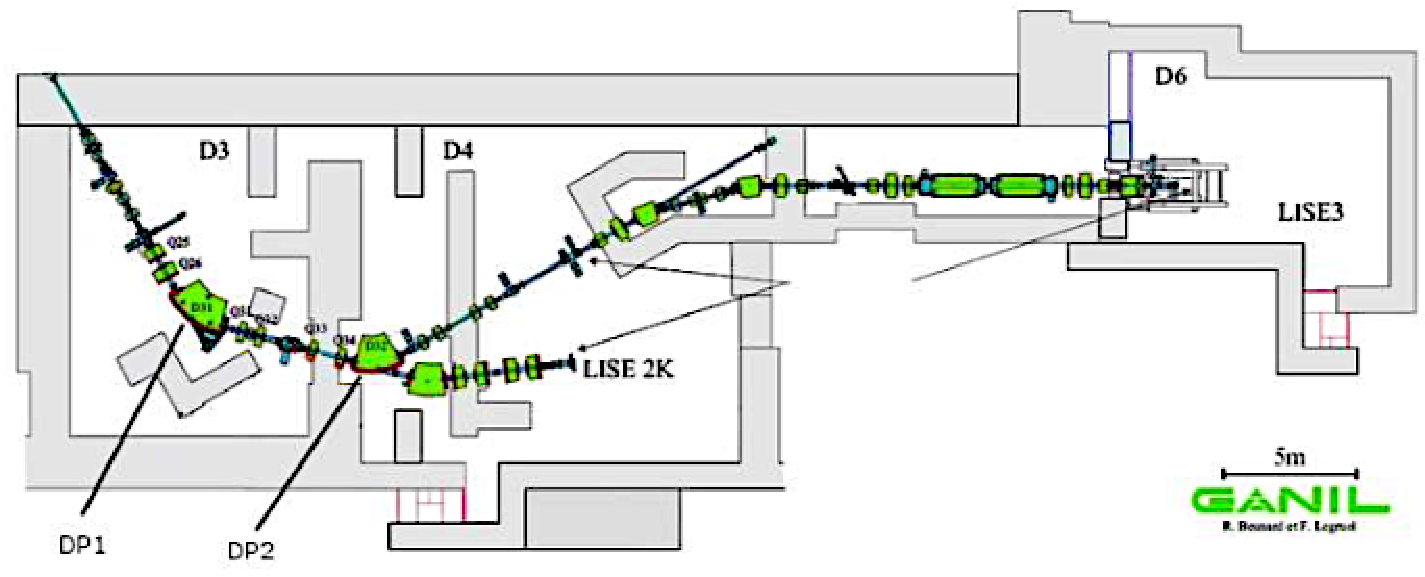}
  \caption{Schematic view of the LISE beam line at GANIL, the experiment took place at the end of the spectrometer in the experimental hall D6.}
  \label{LISE}
\end{figure}

LISE (Fig. \ref{LISE}) is a doubly achromatic spectrometer used for the production of radioactive ion beams from the fragmentation of stable beams on a production target. It is capable of delivering a large range of exotic beams \cite{LISERef}. In the first two experiments of the experimental campaign discussed in this proceeding, radioactive beams were produced from the fragmentation of stable $^{50}$Cr and $^{70}$Zn beams on a Be target, producing respectively a $^{48}$Cr beam at 30 MeV/u for the first experiment and for the second experiment a $^{68}$Ni beam first at 40 MeV/u then later slowed down to 18 MeV/u.

\section{Experimental set-up}

\begin{figure}[ht]
  \centering
  \includegraphics[width=1\textwidth]{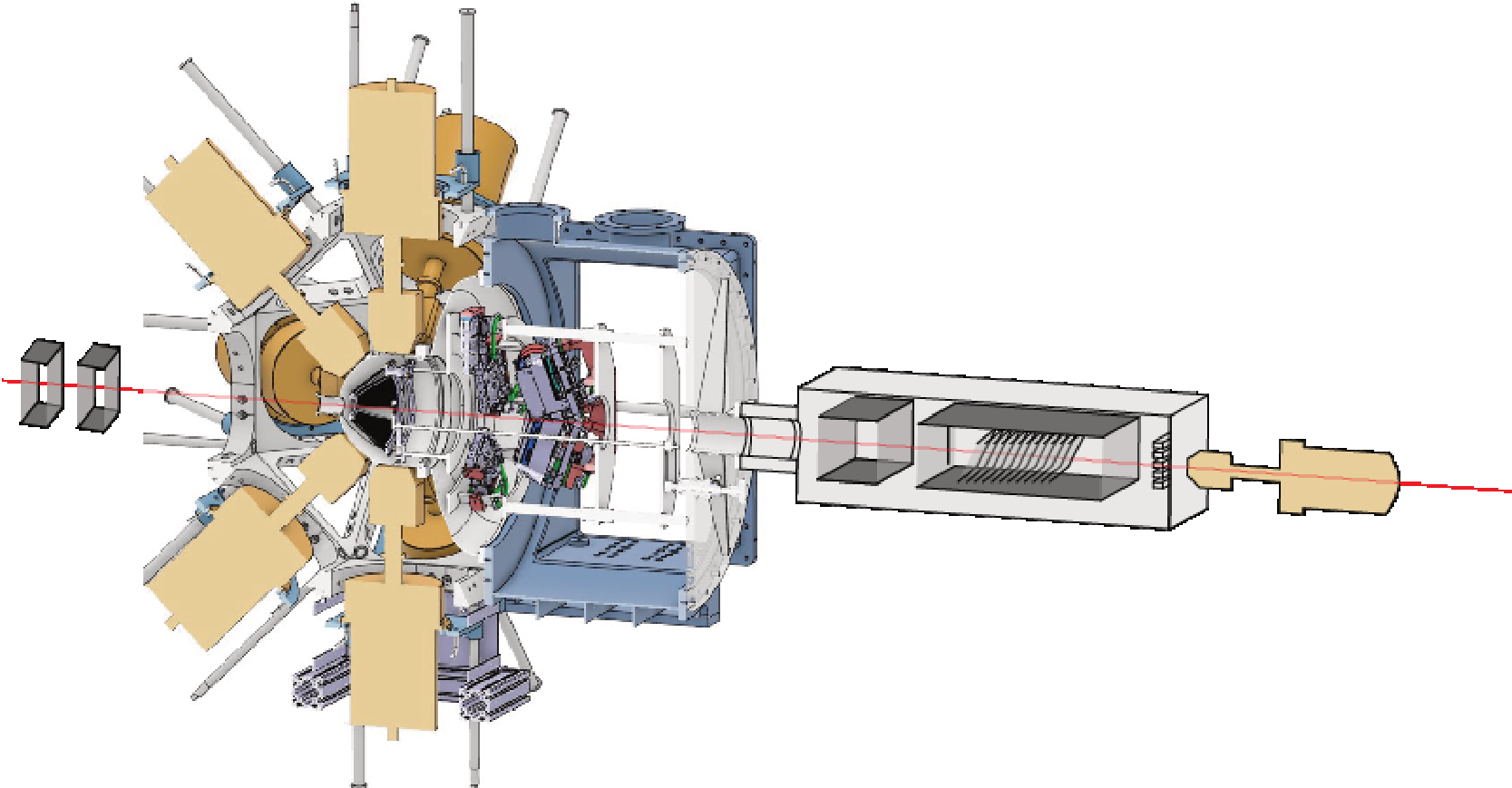}
  \caption{Schematic view of the experimental set-up. The first two grey blocks correspond to the CATS beam tracker, the yellow shape around the target are the EXOGAM clovers, the target is placed at the center of the EXOGAM clovers and is surrounded by the MUGAST detectors and electronics, the ZDD is placed after the reaction/MUGAST chamber.}
  \label{setup}
\end{figure}

The measurement of transfer reactions in inverse kinematics coupled to high resolution $\gamma$-ray spectroscopy requires the development of light-particles detection arrays that combines particle identification, good energy/angular resolution of light ejectiles and excellent $\gamma$-ray transparency. Highly segmented silicon arrays are the best candidates to fulfil these requirements, but previous iteration of these arrays used at GANIL such as for example the MUST2 telescopes \cite{MUST2} had very poor  $\gamma$-ray transparency due mainly to it's integrated electronics. Also to greatly improve $\gamma$-ray detection efficiency, these Si arrays should be made as compact as possible to fit into high resolution $\gamma$-ray spectrometer as for example in the 23 cm radius sphere of AGATA. To overcome these challenges of $\gamma$-ray transparency and efficiency, a new silicon array called GRIT \cite{GRIT} is currently under development. As a first step the MUGAST array combines the MUST2 detectors at forward angles, where no $\gamma$-ray detection is placed, with new trapezoidal shape silicon detectors at backward angles with excellent $\gamma$-ray transparency. For this LISE campaign the MUGAST configuration and mechanics have been redesigned to fit the EXOGAM array and to be suited to the use of fragmentation beams, Fig. \ref{setup} shows a scheme of the full experimental set-up.

\begin{figure}[h]
  \centering
  \includegraphics[width=1\textwidth]{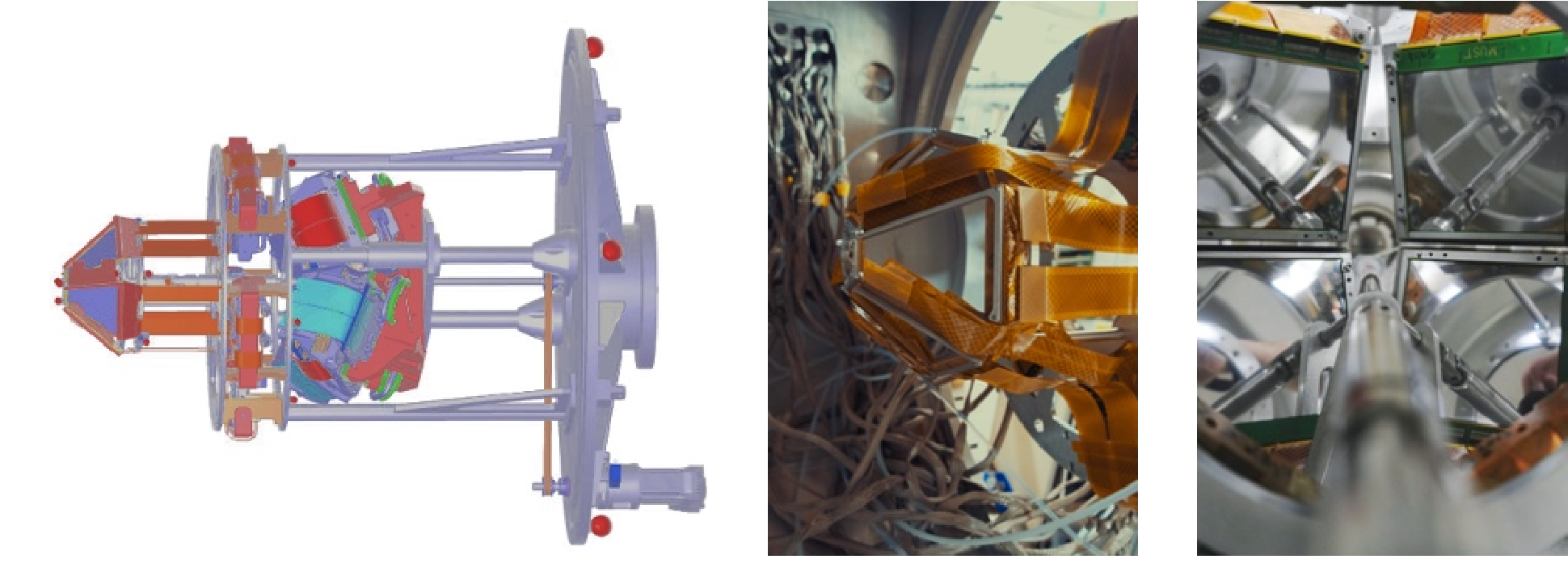}
  \caption{Scheme of the MUGAST detectors inside the chamber on the left, a picture of the trapezoidal DSSD placed at backward angles are visible in the middle while the face of the MUST2 detectors placed at forward angles are visible on the right.}
  \label{MUSTANDMUGAST}
\end{figure}

Fragmented radioactive beams can have large dispersion in position leading to a rather large beam spot (few cm FWHM) on the target, to compensate for this effect, two CATS beam trackers \cite{CATS} have been placed upstream of the target for an event by event reconstruction of the incident beam position. These trackers were also used as a time reference due to their excellent time resolution (of the order of few ps) and for cross-section normalization. The CATS are low pressure multi-wire proportional chambers filled with an isobutane gas (C$_4$H$_{10}$) with a pressure of 8 mbar. The gas is trapped between two 1.5 $\mu$m Mylar foils. Each detector is made of a plane of 71 anodes wires between two segmented cathode containing each 28 strips placed horizontally for one and vertically for the other. 

The MUGAST array (Fig. \ref{MUSTANDMUGAST}) \cite{MUGASTAGATAVAMOS} is a highly segmented silicon array developed for the detection of light particles. It combines five 500 $\mu$m trapezoidal double sided silicon strip detectors (DSSD) with 128x128 strips at backward angles to four MUST2 telescopes at forward angles. The MUST2 telescopes \cite{MUST2} are themselves composed of two stages, a square 300 $\mu$m DSSD with 128x128 strips and a 4x4 CsI crystals stage. The array is capable of measuring the energy and angle of light particles (up to Li) as well as perform time-of-flight ({\it{tof}}) and $\Delta$E-E identification, only at forward angles for the latter. Particular attention has been paid to the array's transparency to $\gamma$-rays to make it compatible with the use of high-purity germanium spectrometer such as EXOGAM or AGATA. 

\begin{figure}
  \centering
  \includegraphics[width=0.5\textwidth]{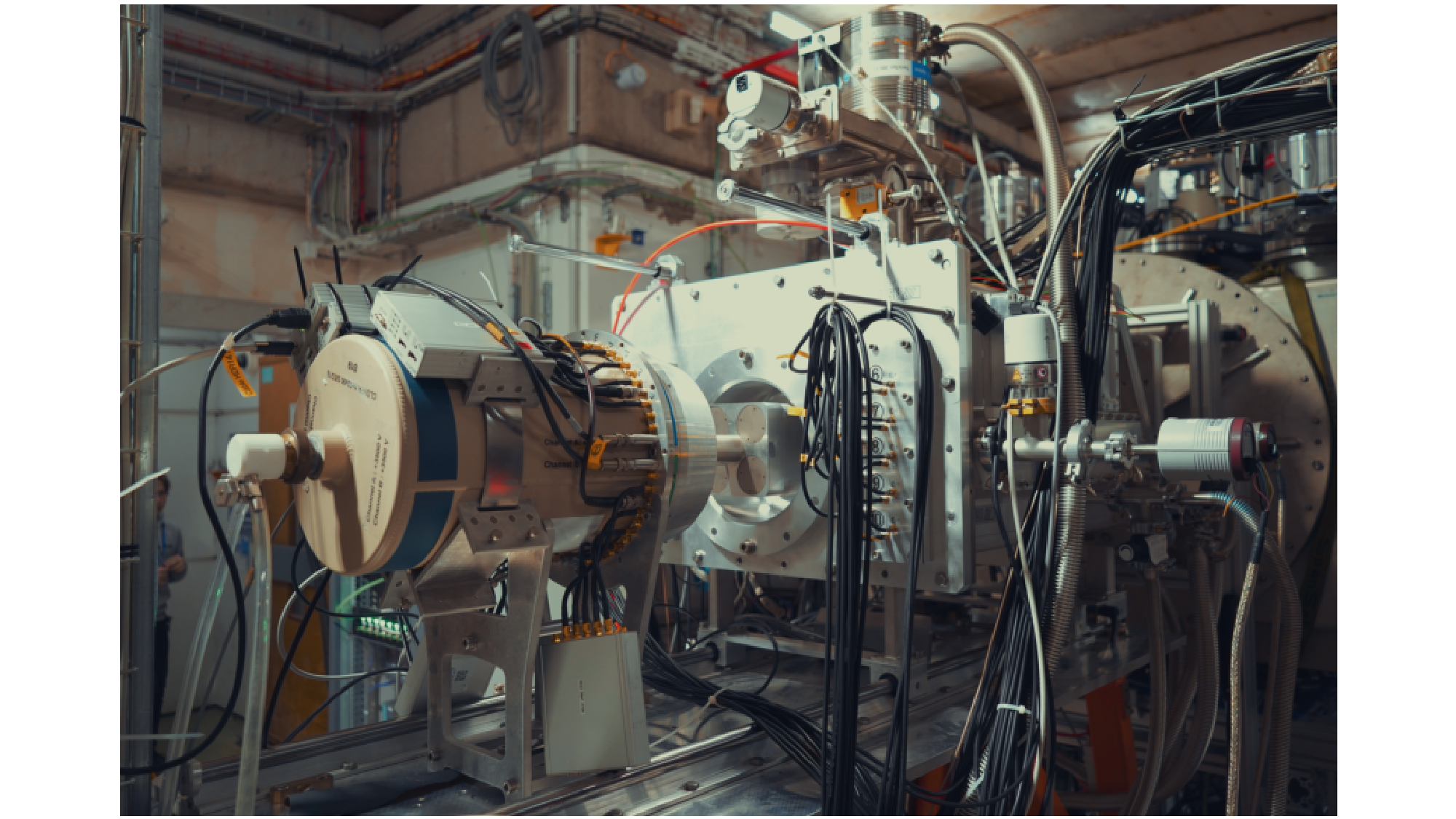}
  \caption{This picture shows the new ZDD of LISE as well as an EXOGAM clover placed at 0$^{\circ}$. The other EXOGAM clovers surrounding the target are identical to this one.}
  \label{EXOGAMANDZDD}
\end{figure}

Twelve EXOGAM clovers (Fig. \ref{EXOGAMANDZDD}) are placed around the target, 8 at 90$^{\circ}$ and 4 at 45$^{\circ}$. EXOGAM is a $\gamma$-ray spectrometer \cite{EXOGAM} composed of high-purity germanium clovers electrically segmented in 4 for an improved Doppler correction. Each clover has also compton shields, BgO ones on the sides and a CsI one at the back. It also uses two different set of gains for the measurement of low and high energy $\gamma$-rays. An additional EXOGAM clover is also placed at 0$^{\circ}$ for the detection and identification of isomers.

For the identification of heavier residues, GANIL has developed a new zero degree detection system for LISE (ZDD) (Fig. \ref{EXOGAMANDZDD}) that was commissioned in 2022. It is composed of two drift chambers (DC), a set of 5 ionization chambers (IC), 5 position sensitive plastic detectors as well as an EXOGAM clover. Particle identification can be performed using the $\Delta$E-E technique with the energy-loss in the IC ($\Delta$E) and the residual energy measured in the plastics, or using the $\Delta$E-{\it{tof}} technique using the {\it{tof}} between the CATS and the plastics. 

\begin{figure}
  \centering
  \includegraphics[width=0.45\textwidth]{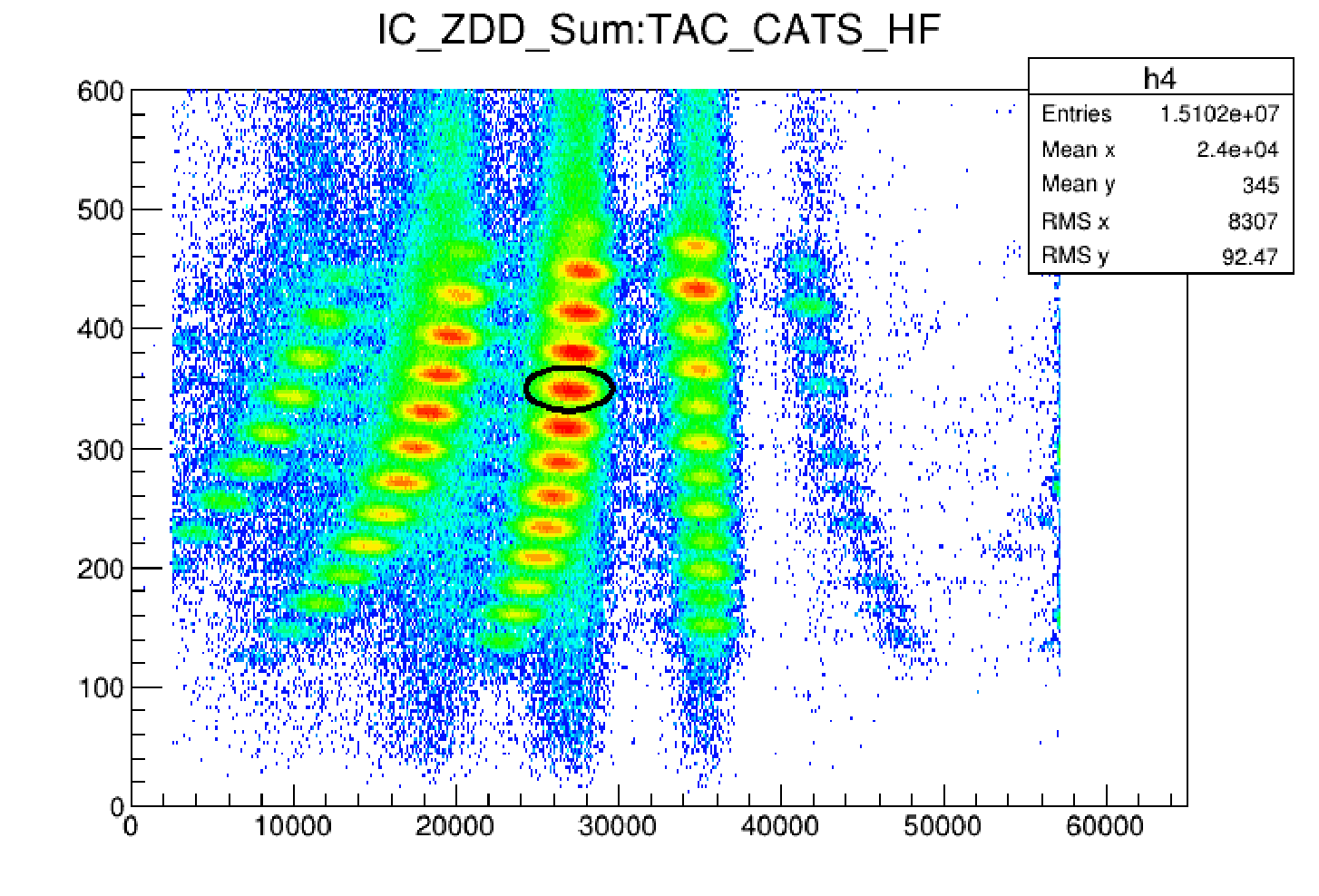}
  \includegraphics[width=0.5\textwidth]{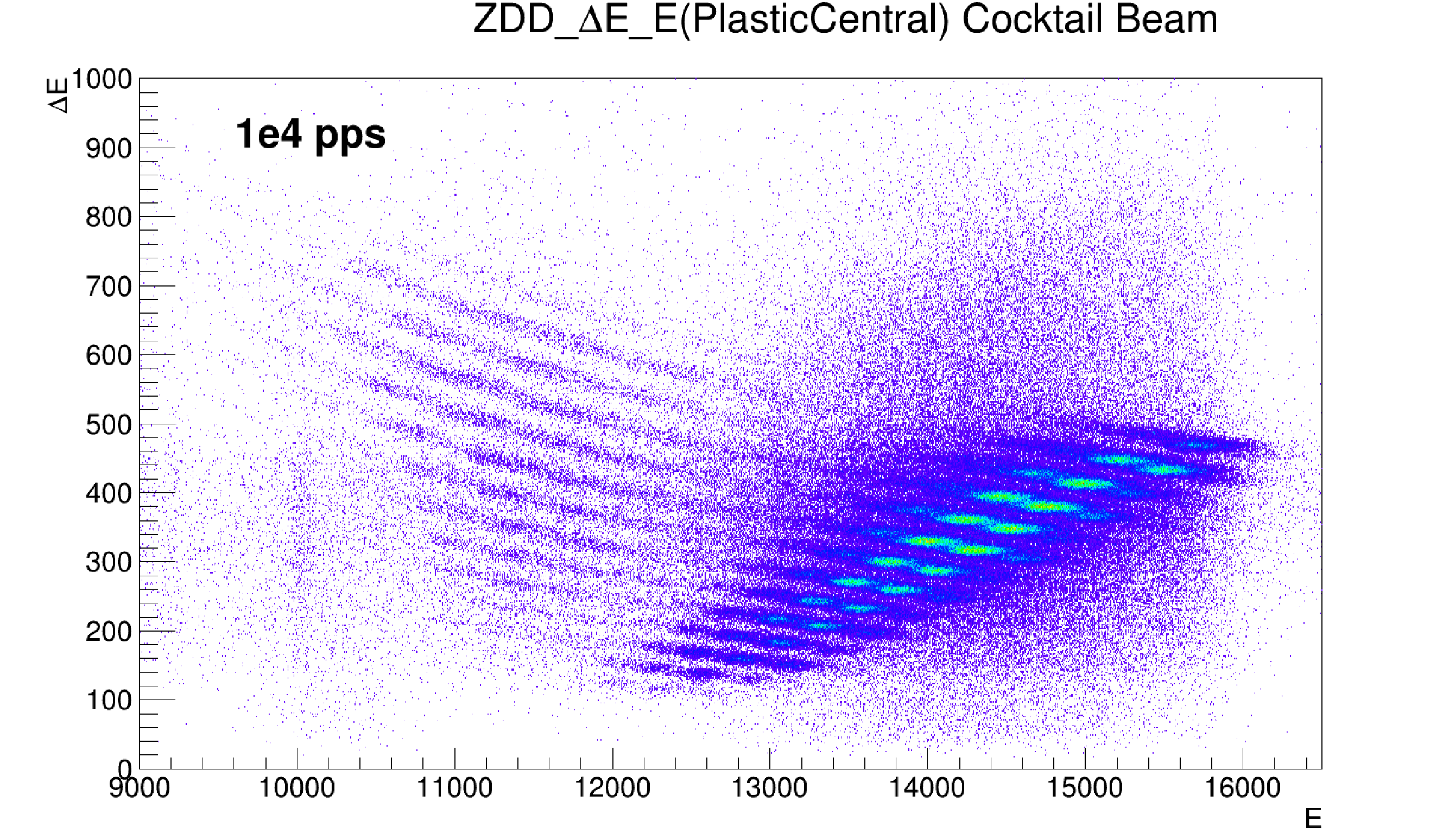}
  \caption{Particle identification of heavy recoils in the ZDD obtained from the fragmentation of a low intensity (1$\times$10$^{4}$ pps) $^{50}$Cr beam at 30 MeV/u. Here there is no selection on the $^{48}$Cr of interest from LISE. $\Delta$E-{\it{tof}} is on the left and $\Delta$E-E on the right. An excellent separation in both mass and charge can be obtained with $\Delta$E-E while the $\Delta$E-{\it{tof}} is limited to mass identification.}
  \label{PIDZDD}
\end{figure}

\section{Performances}

\begin{figure}
  \centering
  \includegraphics[width=0.48\textwidth]{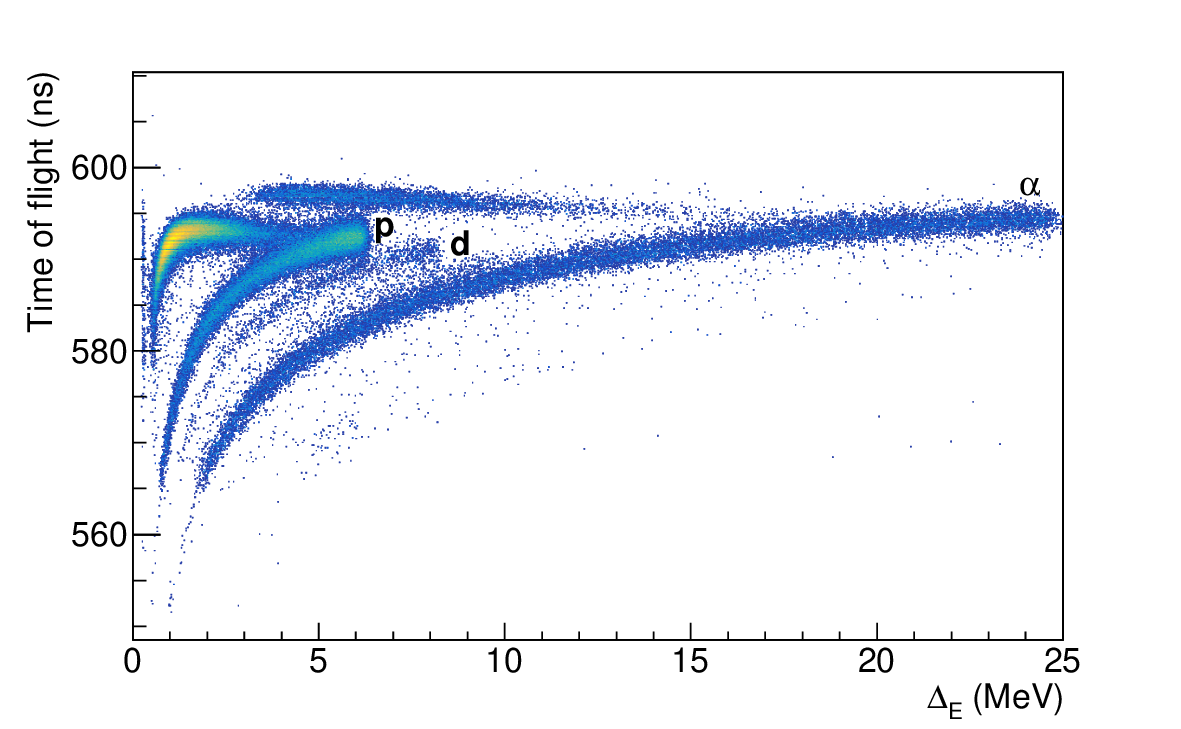}
  \includegraphics[width=0.48\textwidth]{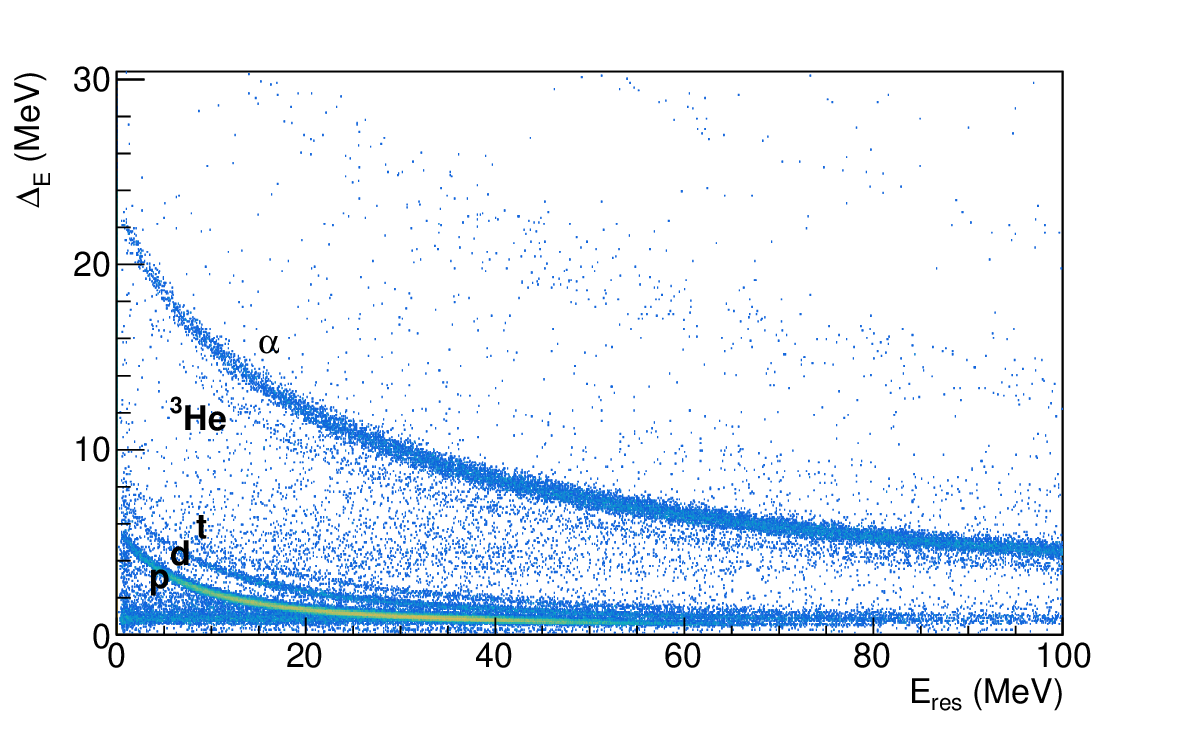}
  \caption{Particle identification of light particles in the MUGAST detectors from $\Delta$E-{\it{tof}} on the left and $\Delta$E-E on the right. The spectra are taken from the forward MUST2 detectors.}
  \label{DeEToF}
\end{figure}

Light particles were identified with the MUGAST array using either energy loss in the DSSD ($\Delta$E) as a function of the {\it{tof}}  between CATS and MUGAST (Fig. \ref{DeEToF}) or in the case of the MUST2 detectors using $\Delta$E as a function of the residual energy in the CsI (E) (Fig. \ref{DeEToF}). The MUGAST detectors covered angles from 5$^{\circ}$ to 30$^{\circ}$ and from 105$^{\circ}$ to 160$^{\circ}$. The DSSD of MUGAST have an energy resolution of $\sim$ 35 keV at $\sim$ 5 MeV determined using a 3 $\alpha$ ($^{239}$Pu $^{241}$Am $^{244}$Cm) source. The final missing mass energy resolution that can be obtained will largely depend on the target thickness and if it is required to use the CsI crystals whose energy resolution can be estimated as follow: $\sigma_{CsI}$ = 0.08$\sqrt{CsI_{Energy}}$~MeV. The $\theta$ and $\phi$ angular resolution of both the forward and backward DSSD is better than 0.5 $^{\circ}$.

\begin{figure}
  \centering
  \includegraphics[width=0.48\textwidth]{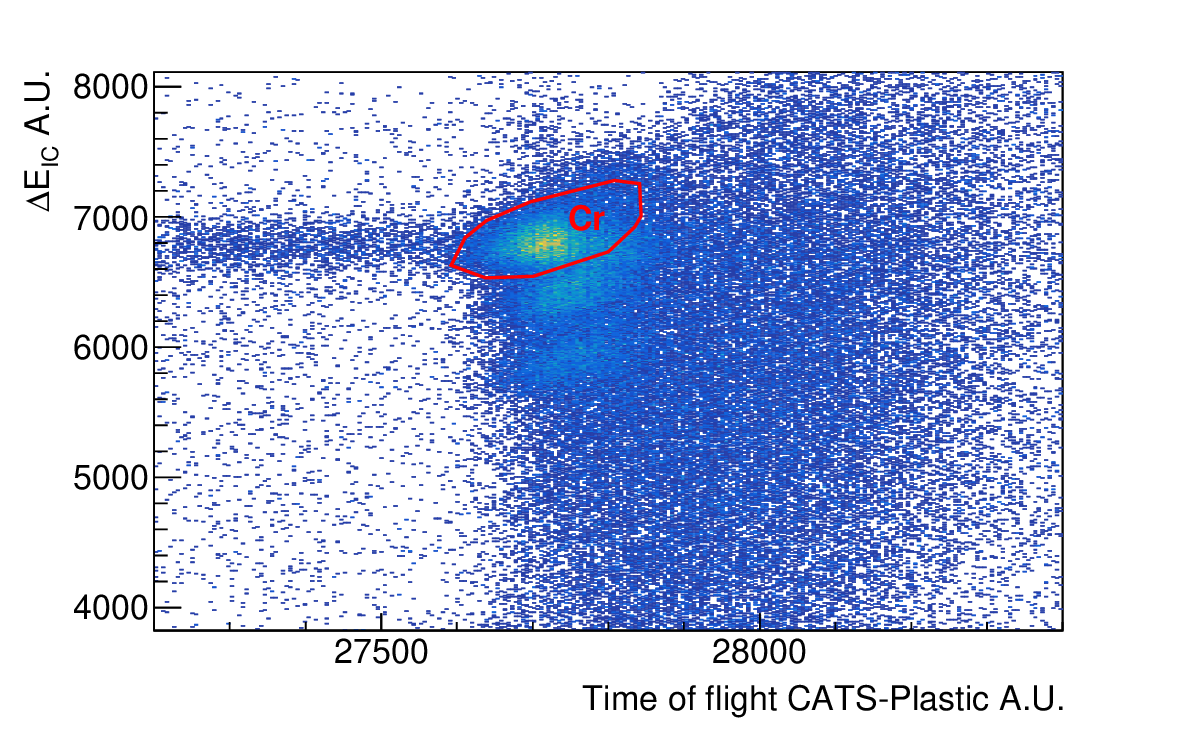}
  \includegraphics[width=0.48\textwidth]{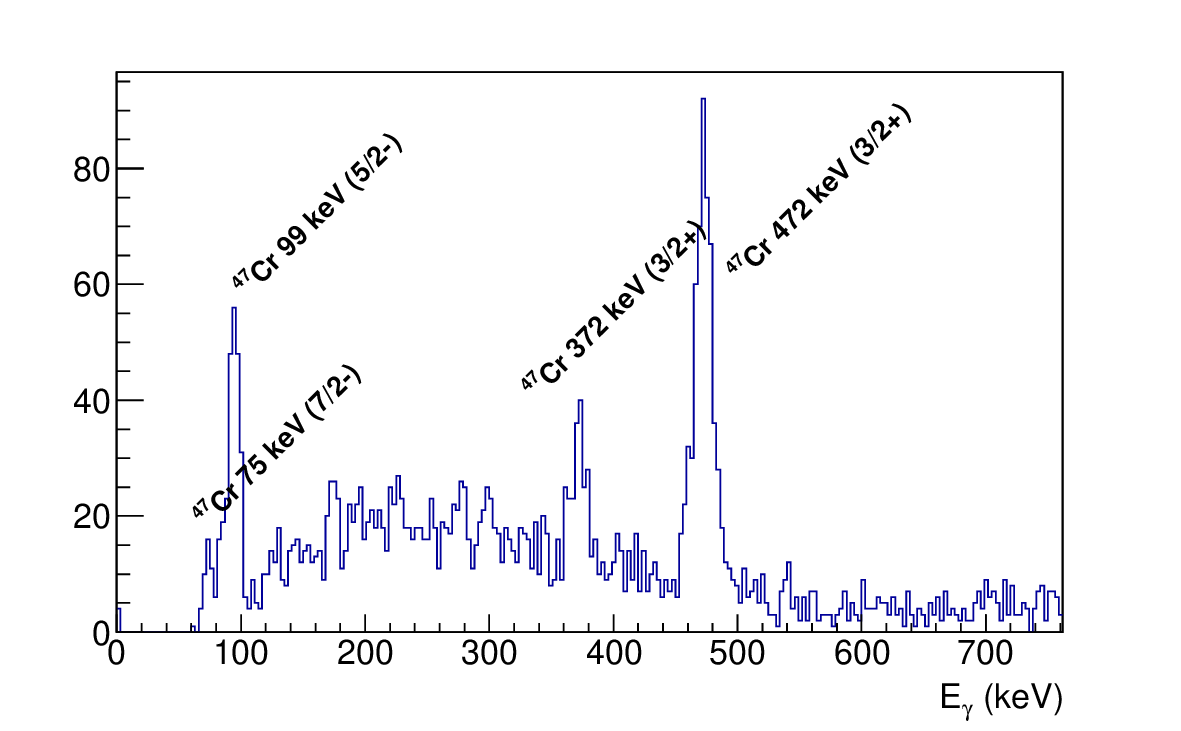}
  \caption{$\Delta$E-{\it{tof}} identification of Cr in ZDD in coincidence with a deuteron in MUST2 on the left, the spectra on right display the $\gamma$ ray in coincidence used to validate that the Cr isotopes corresponds to the $^{47}$Cr of interest.}
  \label{CrID}
\end{figure}


The $\gamma$-rays were detected by 12 EXOGAM clovers. Determination of the photo-peak efficiency using a $^{152}$Eu source demonstrated a 3$\%$ efficiency at 1409 keV, the efficiency is less than the 8 $\%$ expected originally. This is currently under investigation. The Doppler corrected energy resolution at 474 keV obtained online with a beta of 0.24 is 10 keV, it was obtained considering the beam impact position reconstructed event by event. In a more detailed analysis, the energy resolution will be improved by taking into account the trajectories of the light particles detected in the MUGAST detectors.

\begin{figure}[h]
  \centering
  \includegraphics[width=0.5\textwidth]{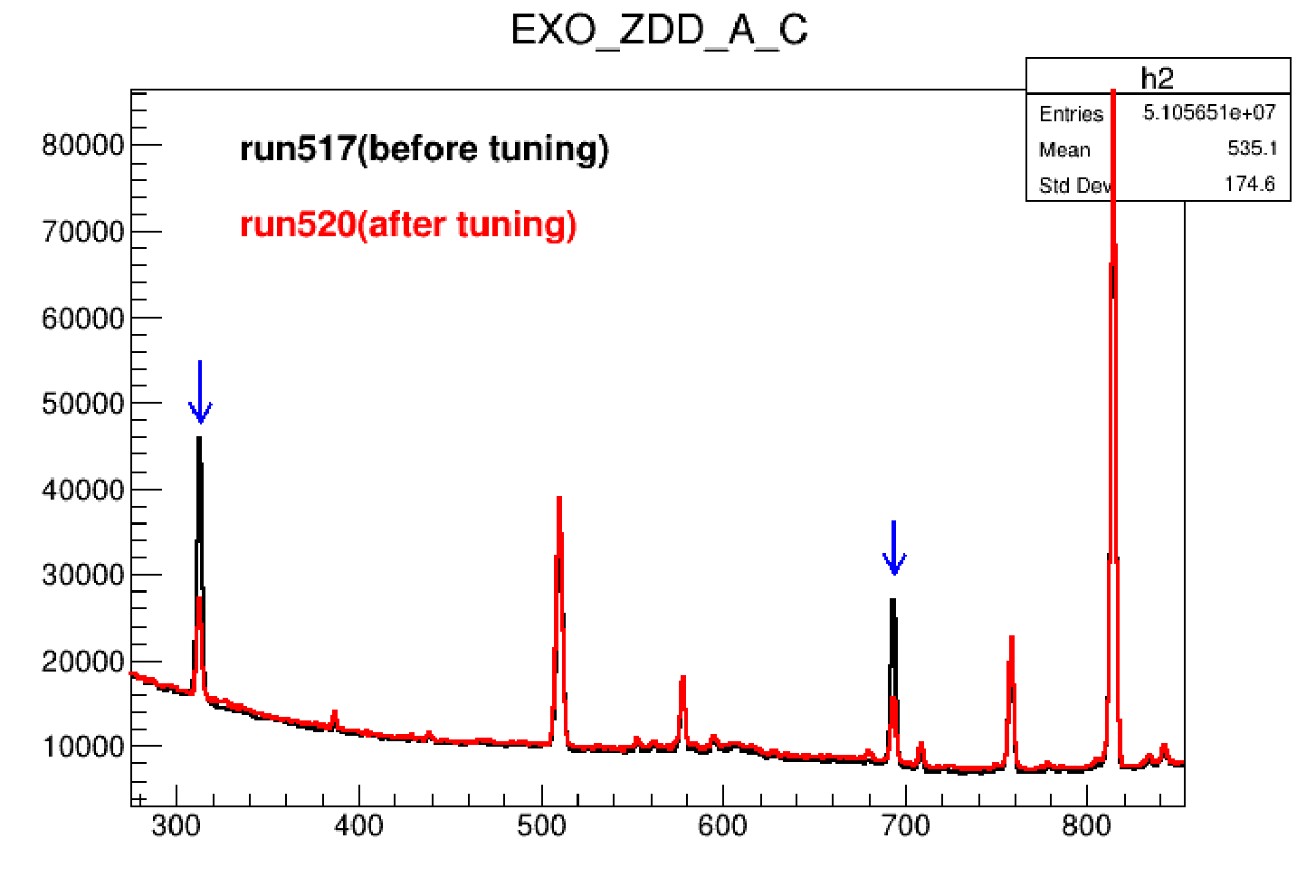}
  \caption{$\gamma$ spectra from the EXOGAM clover at 0$^{\circ}$ before tuning in black and after tuning in red. After tuning, the two typical $\gamma$ transition from the isomer are reduced by 30$\%$.}
  \label{ZDDIsomer}
\end{figure}

The detection of the recoils in the ZDD showed good separation using both the $\Delta$E-E and $\Delta$E-{\it{tof}} techniques. Fig. \ref{PIDZDD} shows an example of such identification at low intensity with a cocktail beam obtained from the fragmentation of $^{50}$Co on the LISE Be target. The maximum acceptable intensity in the ZDD is $\sim$ 2$\times$10$^{5}$ pps after which the large number of pile-up events limits the particle identification. During beam time, due to the degradation of the center plastic detector, only the $\Delta$E-{\it{tof}}  was used. The decay station part of the ZDD also proved to be crucial during beam time as it helped diagnose and limit the isomer contamination of the $^{68}$Ni beam by 30$\%$ as it can be seen on Fig. \ref{ZDDIsomer}.

\begin{figure}
  \centering
  \includegraphics[width=0.48\textwidth]{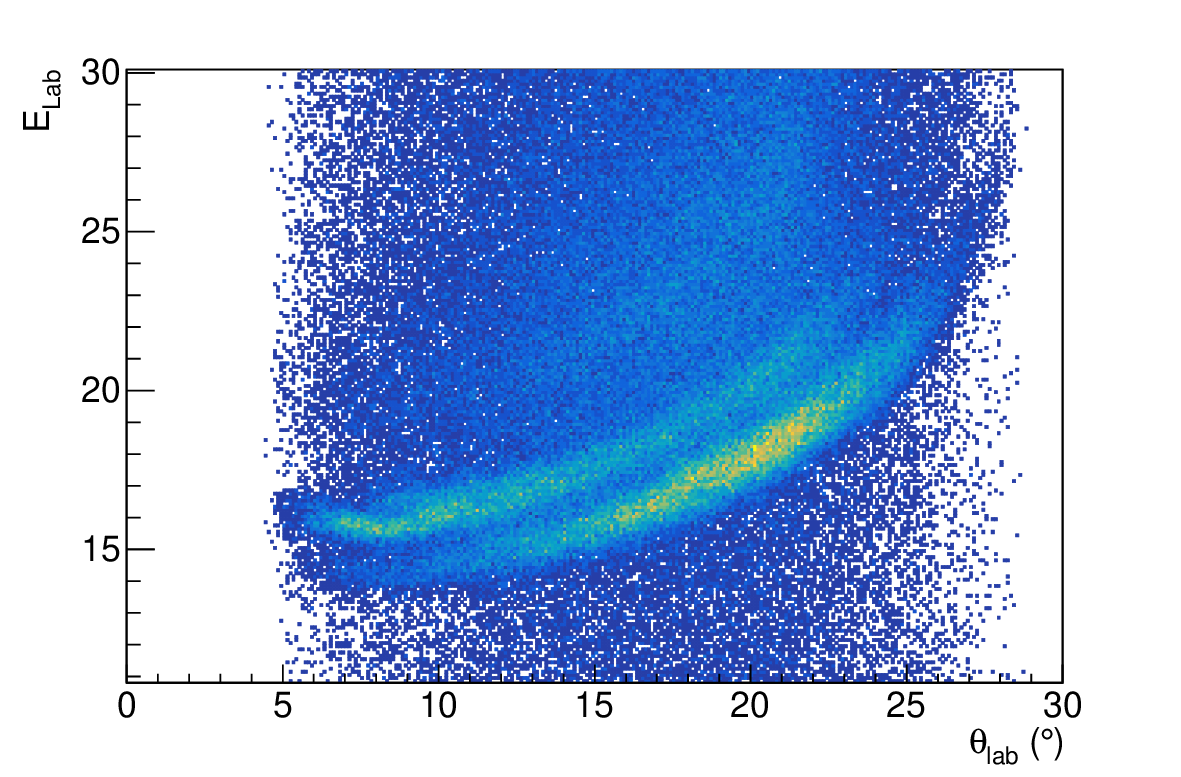}
  \includegraphics[width=0.48\textwidth]{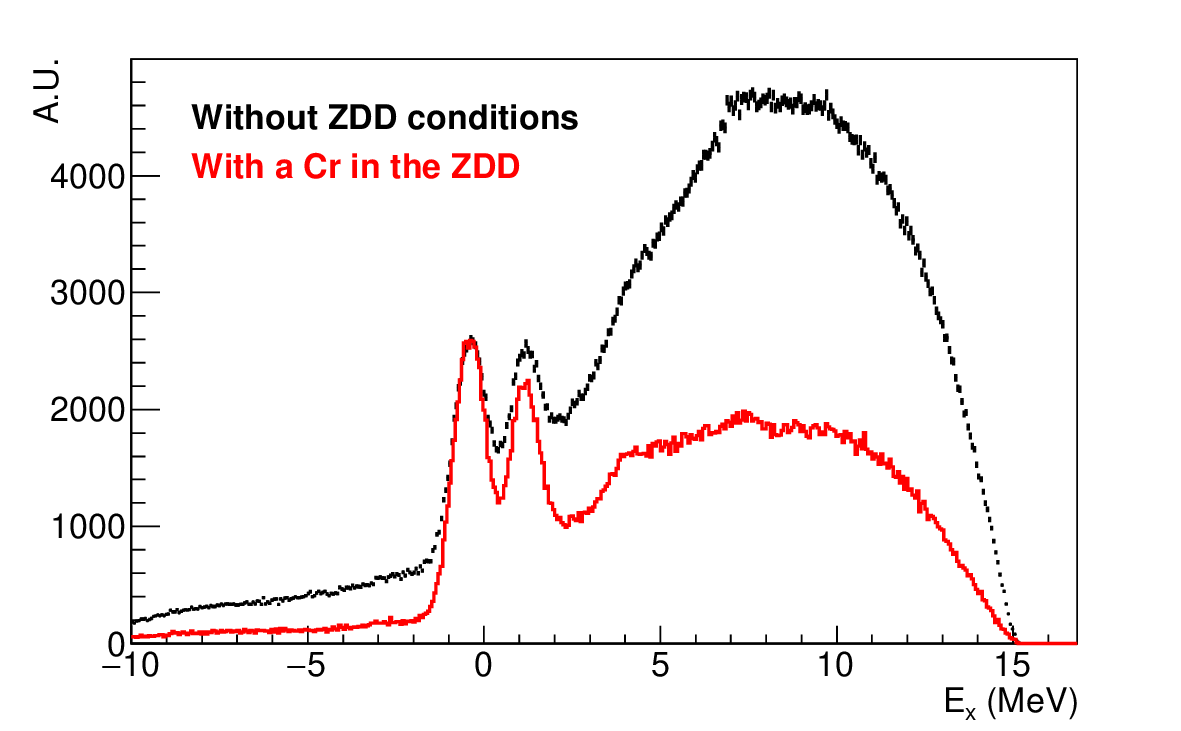}
  \caption{Kinematics of the $^{48}$Cr(p,d)$^{47}$Cr reaction on the left and the corresponding excitation function on the right without conditions on the ZDD in black and with the Cr selection from Fig. \ref{CrID} in red.}
  \label{figKineEx}
\end{figure}

The $^{48}$Cr(p,d)$^{47}$Cr reaction is a good test case to illustrate the performances of the system. Fig. \ref{figKineEx}	shows the excellent background rejection that can be obtained by combining the identification of the deuteron in MUST2 to the detection of Cr isotopes in the ZDD and the identification of $^{47}$Cr $\gamma$-rays in EXOGAM. Fig. \ref{figKineEx} also shows the kinematics and the excitation energy reconstructed using the missing mass technique,  two clear states corresponding to the ground state and an excited state of $^{47}$Cr are clearly visible. The excitation energy resolution obtained using the MUST2 ($\Delta$E + E) measured energy with a 5 mg/cm$^{2}$ target is: 1.3 MeV (FWHM), this online resolution will be further improved with a more detailed analysis, in particular considering a more precise beam reconstruction, improving the CsI calibration and by taking into account the energy loss of the deuteron inside the target. The study of the angular distribution of these states will also make possible the assignment of their spin and parity.

\begin{figure}[h]
  \centering
  \includegraphics[width=0.7\textwidth]{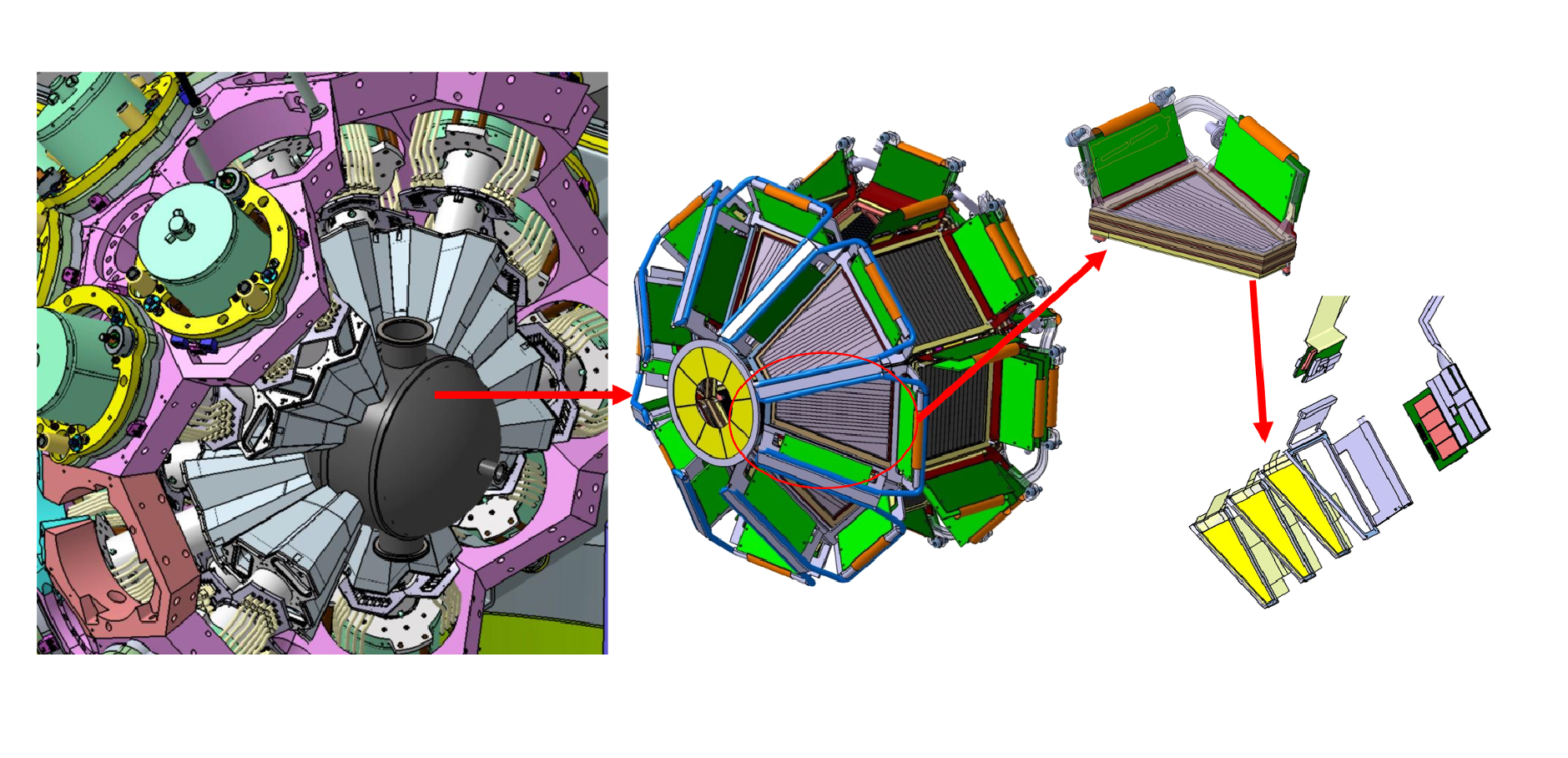}
  \caption{Mechanical drawing of GRIT, on the left the GRIT chamber is seen inside the AGATA $\gamma$-ray spectrometer. In the center the GRIT array which features trapezoidal, square and annular DSSD. On the right an exploded view of the stacked trapezoidal detectors.}
  \label{GRIT}
\end{figure}

\section{Outlooks}

The MUGAST array combined to the detection of heavy recoils and high resolution $\gamma$-ray spectroscopy is a very powerful tool for the study of a wide range of physics topic probed by direct transfer reactions. The success of the previous MUGAST-AGATA-VAMOS campaign from 2019-2021 as well as the one of the two first experiments of the new MUGAST@LISE campaign paves the way for future state-of-the-art measurements at GANIL. 

Four new experiment have already been accepted and should be scheduled in 2024 and 2025 with a broad range of topics such as clustering, astrophysics, shell-model and multi-neutron systems. MUGAST is expected to run at GANIL at least until the end of 2025, after that the new GRIT array will be used in conjunction with the AGATA $\gamma$-ray spectrometer. This new array is designed to fit into AGATA 4$\pi$ and will feature modern electronics, excellent $\gamma$-ray transparency and use a new type of light particles identification via the pulse shape analysis (PSA) technique \cite{PSA}.

\acknowledgments
The authors would like to thank the GANIL staff for their continuous help.

\end{document}